\documentclass[11pt,onecolumn,aps,superscriptaddress,
               amsmath,amssymb,nofootinbib]{revtex4}      
\usepackage{graphicx}

\usepackage{graphicx,epsfig}

\usepackage{dcolumn}
\usepackage{bm}

\newcommand{\Z}{{\mathbb Z}}

\begin{document}

\begin{flushright}
\baselineskip=12pt \normalsize
{ACT-05-09},
{MIFP-09-17}\\
\smallskip
\end{flushright}

\title{Stringy WIMP Detection and Annihilation}

\author{James A. Maxin}
\affiliation{George P. and Cynthia W. Mitchell Institute for
Fundamental Physics, Texas A\&M
University,\\ College Station, TX 77843, USA}
\author{Van E. Mayes}
\affiliation{George P. and Cynthia W. Mitchell Institute for
Fundamental Physics, Texas A\&M
University,\\ College Station, TX 77843, USA}
\author{Dimitri V. Nanopoulos}
\affiliation{George P. and Cynthia W. Mitchell Institute for
Fundamental Physics, Texas A\&M
University,\\ College Station, TX 77843, USA}
\affiliation{Astroparticle Physics Group, Houston
Advanced Research Center (HARC),
Mitchell Campus,
Woodlands, TX~77381, USA; \\
Academy of Athens,
Division of Natural Sciences, 28~Panepistimiou Avenue, Athens 10679,
Greece}

\begin{abstract}
\begin{center}
{\bf ABSTRACT}
\end{center}

We calculate the direct dark matter detection spin-independent and proton spin-dependent cross-sections for a semi-realistic intersecting $D$6-brane model. The cross-sections are compared to the latest constraints of the current dark matter direct detection experiments, as well as the projected results of future dark matter experiments. The allowed parameter space of the intersecting $D$6-brane model is shown with all current experimental constraints, including those regions satisfying the WMAP and Supercritical String Cosmology (SSC) limits on the dark matter density in the universe. Additionally, we compute the indirect detection gamma-ray flux resulting from neutralino annihilation for the $D$6-brane model and compare the flux to the projected sensitivity of the Fermi Gamma-ray Space Telescope. Finally, we compute the direct and indirect detection cross-sections as well as the gamma-ray flux resulting from WIMP annihilations for the one-parameter model for comparison, where the one-parameter model is a highly constrained subset of the mSUGRA parameter space such that the soft supersymmetry breaking terms are functions of the common gaugino mass, which is common to many  string compactifications.  

\end{abstract}

\maketitle

\newpage
\section{Introduction}

Observations in cosmology and astrophysics suggest the presence of a stable dark matter particle. Supersymmetry (SUSY) supplies a satisfactory candidate for a dark matter particle, where R-parity is conserved and the lightest supersymmetric particle (LSP) is stable~\cite{Ellis:1983ew}, which is usually the lightest neutralino $\widetilde{\chi}_{1}^{0}$~\cite{Ellis:1983ew, Goldberg:1983nd}. Two proposed methods of discovering this weakly interacting massive particle (WIMP) are directly through WIMP interactions with ordinary matter and indirectly via the products of WIMP annihilations. The direct detection method searches for elastic scattering of WIMPs off nuclei in underground experiments. The experiments are conducted in deep underground laboratories in an effort to reduce the background to minimal levels. The indirect detection method seeks out debris resulting from WIMP annihilations in the galactic halo. One galactic process that could produce gamma-rays from WIMP annihilation is the process $\widetilde{\chi}_{1}^{0} \widetilde{\chi}^{0}_{1} \rightarrow \gamma \gamma$, where two gamma-rays are produced directly from a WIMP annihilation, and another is $\widetilde{\chi}_{1}^{0} \widetilde{\chi}_{1}^{0} \rightarrow q \overline{q} \rightarrow \pi^{0} \rightarrow \gamma\gamma$. Analyses of direct detection cross-sections and gamma-ray flux within mSUGRA (or CMSSM) models have been completed~\cite{Baltz:2004aw,Ellis:2005mb,Roszkowski:2007fd,Baer:2004qq}. It is, however, a worthwhile pursuit to analyze the direct and indirect detection parameters in alternative models.  

The last few years have seen a great deal of interest in Type II string compactifications. Indeed, intersecting D-brane models (see~\cite{Blumenhagen:2005mu} 
and~\cite{Blumenhagen:2006ci} for reviews)
on Type II orientifolds have become particularly attractive.  In contrast to the standard framework, mSUGRA, the supersymmetry-breaking soft terms
for intersecting D-brane models are in general non-universal~\cite{Kors:2003wf}. Despite substantial progress in constructing such models, most supersymmetric D-brane models suffer from two significant problems. One problem is the lack of gauge coupling unification at the string scale, and the other is the rank one problem in the Standard Model (SM) fermion Yukawa matrices which prevents the generation of mass for the first two generations of quarks and leptons. Nevertheless, there is known one example of an intersecting $D$6-brane model constructed in Type IIA theory on the $T^{6}/(\Z_{2} \times \Z_{2})$ orientifold where these problems have been resolved~\cite{Cvetic:2004ui}~\cite{Chen:2006gd}. This model exhibits automatic gauge coupling unification at the tree-level and it is also possible to obtain the correct Yukawa mass matrices for both up and down-type quarks and leptons for specific values of the moduli VEVs~\cite{Chen:2007px}~\cite{Chen:2007zu}.  Furthermore, the soft-supersymmetry breaking terms for this model have been calculated, where regions in the parameter space were discovered that generate the observed dark matter density and satisfy current experimental constraints~\cite{Chen:2007px}~\cite{Chen:2007zu}. Although this model has many appealing phenomenological features, one issue still to be completely resolved is that of moduli stabilization. This issue has been addressed to an extent~\cite{Chen:2006gd}~\cite{Chen:2007af} by turning on fluxes, but there is still the task of stabilizing the open-string moduli associated with D-brane positions in the internal space and Wilson lines. Only once the 
moduli stabilization issue has been completely addressed can this model be considered fully realistic.  

In this work we show the parameter space allowed by all the experimental constraints for this intersecting $D$6-brane model for varying cases of gravitino masses and tan$\beta$. The spin-independent cross-sections are computed and plotted against the current dark matter detection experiment constraints. Furthermore, we present the proton spin-dependent cross-sections, whereas the computed neutron spin-dependent cross-sections only vary slightly from those of the proton, so the neutron cross-sections are not shown. The gamma-ray flux resulting from neutralino annihilations in the galactic halo for the $D$6-brane model is plotted and compared to the most recent telescope measurements. Finally, in order to compare our results with a model with universal soft terms representing the opposite extreme,  we calculate the spin-independent and spin-dependent cross-sections for the so-called one-parameter model~\cite{Lopez:1993rm,Lopez:1994fz,Lopez:1995hg,Maxin:2008kp}, including the gamma-ray flux. The one-parameter model is a highly constrained small subset of the mSUGRA parameter space such that the soft supersymmetry breaking terms are all functions of the common gaugino mass. In no-scale supergravity models, generically $m_{0} = m_{0}(m_{1/2})$ and $A = A(m_{1/2})$, thus the number of free parameters is reduced to two, $m_{1/2}$ and tan$\beta$. Adopting a strict no-scale framework, one can also fix the $B$-parameter as $B=B(m_{1/2})$, and hence we are led to a {\it one-parameter} model where all of the soft terms may be fixed in terms of $m_{1/2}$. Therefore, the one-parameter model represents a suitable case with which to compare the intersecting $D$6-brane model with non-universal soft supersymmetry breaking terms.  

\section{Low-Energy Effective Action}

In this section, we give a background discussion for the more technically-minded reader which 
describes the way in which
the supersymmetry breaking soft terms are calculated for intersecting
$D6$-brane models. 
In recent years, intersecting D-brane models have provided
an exciting approach toward constructing semi-realistic vacua. 
To summarize, 
D6-branes (in Type IIA) fill three-dimensional
Minkowski space and wrap 3-cycles in the compactified manifold, with a stack of
$N$ branes having a gauge group $U(N)$ (or $U(N/2)$ in the case
 of $T^6/(\Z_2 \times \Z_2)$) in its world volume. The 3-cycles 
wrapped by the D-branes will in general intersect multiple times in the
internal space, resulting in 
a chiral fermion in the bifundamental representation
localized at the intersection between different stacks. The multiplicity
of such fermions is then given by the number of times the 3-cycles intersect.
Due to orientifolding,
for every stack of D6-branes we must also introduce its orientifold
images.  Thus, the D6-branes may also have intersections with the images of other stacks,
also resulting in fermions in bifundamental representations.
Each stack may also intersect its own images, resulting in chiral fermions 
in the symmetric and antisymmetric representations.  
In addition, there are constraints that must be satisfied for the 
consistency of the model, namely the requirement for Ramond-Ramond
tadpole cancellation and to have a spectrum with $\mathcal{N}=1$
supersymmetry.

To discuss the low-energy phenomenology we start from the low-energy
effective action.  From the effective scalar potential it is
possible to study the stability~\cite{Blumenhagen:2001te}, the
tree-level gauge couplings \cite{CLS1, Shiu:1998pa,
Cremades:2002te}, gauge threshold corrections \cite{Lust:2003ky},
and gauge coupling unification \cite{Antoniadis:Blumen}.  The
effective Yukawa couplings \cite{Cremades:2003qj, Cvetic:2003ch},
matter field K\"ahler metric and soft-SUSY breaking terms have
also been investigated \cite{Kors:2003wf}.  A more detailed
discussion of the K\"ahler metric and string scattering of gauge,
matter, and moduli fields has been performed in
\cite{Lust:2004cx}. Although turning on Type IIB 3-form fluxes can
break supersymmetry from the closed string sector
~\cite{Cascales:2003zp, MS, CL, Cvetic:2005bn, Kumar:2005hf,
Chen:2005cf}, there are additional terms in the superpotential
generated by the fluxes and there is currently no satisfactory
model which incorporates this. Thus, we do not consider this option
in the present work.  In principle, it should be possible to
specify the exact mechanism by which supersymmetry is broken, and
thus to make very specific predictions.  However, for the present
work, we will adopt a parametrization of the SUSY breaking so that
we can study it generically.

The $\mathcal{N}=1$ supergravity action depends upon three
functions, the holomorphic gauge kinetic function, $f$, K\a"ahler
potential $K$, and the superpotential $W$.  Each of these will in
turn depend upon the moduli fields which describe the background
upon which the model is constructed. The holomorphic gauge kinetic
function for a D6-brane wrapping a calibrated three-cycle $\Pi$ is given
by (see \cite{Blumenhagen:2006ci} for a detailed discussion and explanation of the notation)
\begin{equation}
f_P = \frac{1}{2\pi \ell_s^3}\left[e^{-\phi}\int_{\Pi_P} \mbox{Re}(e^{-i\theta_P}\Omega_3)-i\int_{\Pi_P}C_3\right].
\end{equation}
In terms of the three-cycle wrapped by the stack of branes, we have
\begin{equation}
\int_{\Pi_a}\Omega_3 = \frac{1}{4}\prod_{i=1}^3(n_a^iR_1^i + 2^{-\beta_i}il_a^iR_2^i).
\end{equation}
where $n^i$ and $l^i$ are the wrapping numbers of the D-branes on the $i$th two-torus, from which it follows that
\begin{eqnarray}
f_P &=&
\frac{1}{4\kappa_P}(n_P^1\,n_P^2\,n_P^3\,s-\frac{n_P^1\,l_P^2\,l_P^3\,u^1}{2^{(\beta_2+\beta_3)}}-\frac{n_P^2\,l_P^1\,l_P^3\,u^2}{2^{(\beta_1+\beta_3)}}-
\frac{n_P^3\,l_P^1\,l_P^2\,u^3}{2^{(\beta_1+\beta_2)}}),
\label{kingauagefun}
\end{eqnarray}
where $\kappa_P = 1$ for $SU(N_P)$ and $\kappa_P = 2$ for
$USp(2N_P)$ or $SO(2N_P)$ gauge groups and where we use the $s$ and
$u$ moduli in the supergravity basis.  In the string theory basis,
we have the dilaton $S$, three K\"ahler moduli $T^i$, and three
complex structure moduli $U^i$~\cite{Lust:2004cx}. These are related to the
corresponding moduli in the supergravity basis by
\begin{eqnarray}
\mathrm{Re}\,(s)& =&
\frac{e^{-{\phi}_4}}{2\pi}\,\left(\frac{\sqrt{\mathrm{Im}\,U^{1}\,
\mathrm{Im}\,U^{2}\,\mathrm{Im}\,U^3}}{|U^1U^2U^3|}\right)
\nonumber \\
\mathrm{Re}\,(u^j)& =&
\frac{e^{-{\phi}_4}}{2\pi}\left(\sqrt{\frac{\mathrm{Im}\,U^{j}}
{\mathrm{Im}\,U^{k}\,\mathrm{Im}\,U^l}}\right)\;
\left|\frac{U^k\,U^l}{U^j}\right| \qquad (j,k,l)=(\overline{1,2,3})
\nonumber \\
\mathrm{Re}(t^j)&=&\frac{i\alpha'}{T^j} \label{idb:eq:moduli}
\end{eqnarray}
and $\phi_4$ is the four-dimensional dilaton.
To second order in the string matter fields, the K\a"ahler potential is given by
\begin{eqnarray}
K(M,\bar{M},C,\bar{C}) = \hat{K}(M,\bar{M}) + \sum_{\mbox{untwisted}~i,j} \tilde{K}_{C_i \bar{C}_j}(M,\bar{M})C_i \bar{C}_j + \\ \nonumber \sum_{\mbox{twisted}~\theta} \tilde{K}_{C_{\theta} \bar{C}_{\theta}}(M,\bar{M})C_{\theta}\bar{C}_\theta.
\end{eqnarray}
The untwisted moduli $C_i$, $\bar{C}_j$ are light, non-chiral
scalars from the field theory point of view, associated with the
D-brane positions and Wilson lines.  These fields are not observed
in the MSSM, and if present in the low energy spectra they
may disrupt the gauge coupling unification.  Clearly, these
fields must get a large mass through some mechanism.  One way to
accomplish this is to require the D-branes to wrap rigid cycles,
which freezes the open string moduli~\cite{Blumenhagen:2005tn}.

For twisted moduli arising from strings stretching between stacks
$P$ and $Q$, we have $\sum_j\theta^j_{PQ}=0$, where $\theta^j_{PQ} =
\theta^j_Q - \theta^j_P$ is the angle between the cycles wrapped
by the stacks of branes $P$ and $Q$ on the $j^{th}$ torus
respectively. Then, for the K\a"ahler metric in Type IIA theory we find
the following two cases:

\begin{itemize}

\item $\theta^j_{PQ}<0$, $\theta^k_{PQ}>0$, $\theta^l_{PQ}>0$

\begin{eqnarray}
\tilde{K}_{PQ} &=& e^{\phi_4} e^{\gamma_E (2-\sum_{j = 1}^3
\theta^j_{PQ}) }
\sqrt{\frac{\Gamma(\theta^j_{PQ})}{\Gamma(1+\theta^j_{PQ})}}
\sqrt{\frac{\Gamma(1-\theta^k_{PQ})}{\Gamma(\theta^k_{PQ})}}
\sqrt{\frac{\Gamma(1-\theta^l_{PQ})}{\Gamma(\theta^l_{PQ})}}
\nonumber \\ && (t^j + \bar{t}^j)^{\theta^j_{PQ}} (t^k +
\bar{t}^k)^{-1+\theta^k_{PQ}} (t^l +
\bar{t}^l)^{-1+\theta^l_{PQ}}.
\end{eqnarray}

\item $\theta^j_{PQ}<0$, $\theta^k_{PQ}<0$, $\theta^l_{PQ}>0$

\begin{eqnarray}
\tilde{K}_{PQ} &=& e^{\phi_4} e^{\gamma_E (2+\sum_{j = 1}^3
\theta^j_{PQ}) }
\sqrt{\frac{\Gamma(1+\theta^j_{PQ})}{\Gamma(-\theta^j_{PQ})}}
\sqrt{\frac{\Gamma(1+\theta^k_{PQ})}{\Gamma(-\theta^k_{PQ})}}
\sqrt{\frac{\Gamma(\theta^l_{PQ})}{\Gamma(1-\theta^l_{PQ})}}
\nonumber \\ && (t^j + \bar{t}^j)^{-1-\theta^j_{PQ}} (t^k +
\bar{t}^k)^{-1-\theta^k_{PQ}} (t^l + \bar{t}^l)^{-\theta^l_{PQ}}.
\end{eqnarray}

\end{itemize}

For branes which are parallel on at least one torus, giving rise
to non-chiral matter in bifundamental representations (for example,
the Higgs doublets), the K\a"ahler metric is
\begin{equation}
\hat{K}=((s+\bar{s})(t^1+\bar{t}^1)(t^2+\bar{t}^2)(u^3+\bar{u}^3))^{-1/2}.
\label{nonchiralK}
\end{equation}
The superpotential is given by
\begin{equation}
W = \hat{W}+ \frac{1}{2}\mu_{\alpha\beta}(M)C^{\alpha}C^{\beta} + \frac{1}{6}Y_{\alpha\beta\gamma}(M)C^{\alpha\beta\gamma}+\cdots
\end{equation}
while the minimum of the F part of the tree-level supergravity
scalar potential $V$ is given by
\begin{equation}
V(M,\bar{M}) = e^G(G_M K^{MN} G_N -3) = (F^N K_{NM} F^M-3e^G),
\end{equation}
where
$G_M=\partial_M G$ and $K_{NM}=\partial_N \partial_M K$, $K^{MN}$
is inverse of $K_{NM}$, and the auxiliary fields $F^M$ are given
by
\begin{equation}
F^M=e^{G/2} K^{ML}G_L. \label{aux}
\end{equation}
Supersymmetry is broken when some of the F-terms of the hidden sector fields $M$
acquire VEVs. This then results in soft terms being generated in
the observable sector. For simplicity, it is assumed in this
analysis that the $D$-term does not contribute (see
\cite{Kawamura:1996ex}) to the SUSY breaking.  Then the goldstino
is absorbed by the gravitino via the superHiggs effect. The
gravitino then obtains a mass
\begin{equation}
m_{3/2}=e^{G/2},
\end{equation}
which we will take to be 500 GeV and 700 GeV in the following. The
normalized gaugino mass parameters, scalar mass-squared
parameters, and trilinear parameters respectively may be given in
terms of the K\a"ahler potential, the gauge kinetic function, and
the superpotential as
\begin{eqnarray}
M_P &=& \frac{1}{2\mbox{Re}f_P}(F^M\partial_M f_P), \\ \nonumber
m^2_{PQ} &=& (m^2_{3/2} + V_0) - \sum_{M,N}\bar{F}^{\bar{M}}F^N\partial_{\bar{M}}\partial_{N}log(\tilde{K}_{PQ}), \\ \nonumber
A_{PQR} &=& F^M\left[\hat{K}_M + \partial_M log(Y_{PQR}) - \partial_M log(\tilde{K}_{PQ}\tilde{K}_{QR}\tilde{K}_{RP})\right],
\label{softterms}
\end{eqnarray}
where $\hat{K}_M$ is the K\a"ahler metric appropriate for branes
which are parallel on at least one torus, i.e. involving
non-chiral matter.  

The above formulas for the soft terms depend on the Yukawa couplings, via the superpotential.  An important consideration is whether or not this should cause any modification to the low-energy spectrum.  However, this turns out not to be the case since the Yukawas in the soft term formulas are not the same as the physical Yukawas, which arise from world-sheet instantons and are proportional to $exp({-A})$, where $A$ is the world-sheet area of the triangles formed by a triplet of intersections at which
the Standard Model fields are localized.  The physical Yukawa couplings in Type IIA depend on the K\a"ahler moduli and the open-string moduli.  This ensures that the Yukawa couplings present in the soft terms do not depend on either the complex-structure moduli or dilaton (in the supergravity basis).  Thus, the
Yukawa couplings will not affect the low-energy spectrum in the case of $u$-moduli dominant and mixed $u$ and $s$ dominant supersymmetry breaking.

To determine the SUSY soft breaking
parameters, and therefore the spectra of the models, we introduce
the VEVs of the auxiliary fields Eq. (\ref{aux}) for the
dilaton, complex and K\"ahler moduli \cite{Brignole:1993dj}:
\begin{eqnarray}
&& F^s=2\sqrt{3}C m_{3/2} {\rm Re}(s) \Theta_s e^{-i\gamma_s},
\nonumber \\
&&F^{\{u,t\}^i} = 2\sqrt{3}C m_{3/2}( {\rm Re}  ({u}^i) \Theta_i^u
e^{-i\gamma^u_i}+  {\rm Re} ({t}^i) \Theta_i^t
e^{-i\gamma_i^t}).
\end{eqnarray}
The factors $\gamma_s$ and $\gamma_i$ are the CP violating phases of the moduli, while
the constant $C$
is given by
\begin{equation}
C^2 = 1+ \frac{V_0}{3 m^2_{3/2}}.
\end{equation}
The goldstino is absorbed into the gravitino by $\Theta_S$ in $S$
field space, and $\Theta_i$ parameterize the goldstino direction
in $U^i$ space,  where $\sum (|\Theta_i^u|^2 + |\Theta_i^t|^2) + |\Theta_s|^2 =1$. The
goldstino angle $\Theta_s$ determines the degree to which SUSY
breaking is being dominated by the dilaton $s$ and/or complex
structure ($u^i$) and K\"ahler ($t^i$) moduli.  As suggested earlier, we
will not consider the case of $t$-moduli dominant supersymmetry breaking, 
since in this case the soft terms are not independent of the Yukawa couplings.

\section{Parameter Space and Supersymmetry Spectra}

The set of soft terms at the unification scale are generated in the same manner as was performed in~\cite{Chen:2007zu} for $\textit{u}$-moduli dominated SUSY breaking.  The soft terms are then input into {\tt MicrOMEGAs 2.0.7}~\cite{Belanger:2006is} using {\tt SuSpect 2.34}~\cite{Djouadi:2002ze} as a front end to run the soft terms down to the electroweak scale via the Renormalization Group Equations (RGEs) and then to calculate the corresponding relic neutralino density.  We take the top quark mass to be
$m_t = 172.6$~GeV~\cite{:2008vn}, and leave tan~$\beta$ as a free parameter while $\mu$ is determined by the requirement of radiative electroweak symmetry breaking (REWSB). However, we do take $\mu > 0$ as suggested by the results of $g_{\mu}-2$ for the muon. The results are then filtered according to the following criteria: 

\begin{enumerate}

\item The WMAP 5-year data~\cite{Hinshaw:2008kr} for the cold dark matter density,  0.1109 $\leq \Omega_{\chi^o} h^{2} \leq$ 0.1177.  We also consider the WMAP 2$\sigma$ results~\cite{Spergel:2006hy}, 0.095 $\leq \Omega_{\chi^o} h^{2} \leq$ 0.129. In addition, we look at the SSC model~\cite{Antoniadis:1988aa} for the dark matter density, in which a dilution factor of $\cal{O}$(10) is allowed~\cite{Lahanas:2006hf}, where $\Omega_{\chi^o} h^{2} \leq$ 1.1. For a discussion of the SSC model within the context of mSUGRA, see~\cite{Dutta:2008ge}. We investigate two cases, one where a neutralino LSP is the dominant component of the dark matter and another where it makes up a subdominant component such that
0 $\leq \Omega_{\chi^o} h^{2} \leq$ 0.1177, 0 $\leq \Omega_{\chi^o} h^{2} \leq$ 0.129, and 0 $\leq \Omega_{\chi^o} h^{2} \leq$ 1.1. This allows for the possibility that dark matter could be composed of matter such as axions, cryptons, or other particles.

\item The experimental limits on the Flavor Changing Neutral Current (FCNC) process, $b \rightarrow s\gamma$. The results from the Heavy Flavor Averaging Group (HFAG)~\cite{Barberio:2007cr}, in addition to the BABAR, Belle, and CLEO results, are: $Br(b \rightarrow s\gamma) = (355 \pm 24^{+9}_{-10} \pm 3) \times 10^{-6}$. There is also a more recent estimate~\cite{Misiak:2006zs} of $Br(b \rightarrow s\gamma) = (3.15 \pm 0.23) \times 10^{-4}$. For our analysis, we use the limits $2.86 \times 10^{-4} \leq Br(b \rightarrow s\gamma) \leq 4.18 \times 10^{-4}$, where experimental and
theoretical errors are added in quadrature.

\item The anomalous magnetic moment of the muon, $g_{\mu} - 2$. For this analysis we use the 2$\sigma$ level boundaries, $11 \times 10^{-10} < a_{\mu} < 44 \times 10^{-10}$~\cite{Bennett:2004pv}.

\item The process $B_{s}^{0} \rightarrow \mu^+ \mu^-$ where the decay has a $\mbox{tan}^6\beta$ dependence. We take the upper bound to be $Br(B_{s}^{0} \rightarrow \mu^{+}\mu^{-}) < 5.8 \times 10^{-8}$~\cite{:2007kv}.

\item The LEP limit on the lightest CP-even Higgs boson mass, $m_{h} \geq 114$ GeV~\cite{Barate:2003sz}.

\end{enumerate}

A scan of the parameter space allowed by the aforementioned five experimental constraints was performed for various values of the gravitino mass and tan$\beta$, with the goal to determine the range of the gravitino mass where the upper limit is the mass at which SUSY events become observable at the Large Hadron Collider (LHC) above the SM background, and at the lower limit the Higgs mass becomes too light and violates the LEP constraint. We discover the upper limit to be $m_{3/2}$ $\approx$ 700 GeV and the lower limit to be in the range $m_{3/2}$ = 400 $\sim$ 500 GeV. Thus, we calculate the relic density, experimental constraints, and subsequently, the direct detection cross-sections and indirect detection gamma-ray flux for $m_{3/2}$ = 500 GeV and $m_{3/2}$ = 700 GeV. For each of these gravitino masses, the calculations were completed for tan$\beta$ = 10, 25, and 46. Regions of the parameter space satisfying all the experimental constraints exist for five of the six cases; only $m_{3/2}$ = 700 GeV, tan$\beta$ = 10 produced no regions that satisfied the constraints. Additional values of tan$\beta$ were run for $m_{3/2}$ = 700 GeV, though tan$\beta$ = 25 is close to the minimum tan$\beta$ that violates none of the constraints. Thus, we study five cases in this paper: $m_{3/2}$ = 500 GeV and tan$\beta$ = 10, $m_{3/2}$ = 500 GeV and tan$\beta$ = 25, $m_{3/2}$ = 500 GeV and tan$\beta$ = 46, $m_{3/2}$ = 700 GeV and tan$\beta$ = 25, $m_{3/2}$ = 700 GeV and tan$\beta$ = 46. 

\begin{figure}[htp]
	\centering
		\includegraphics[width=0.42\textwidth]{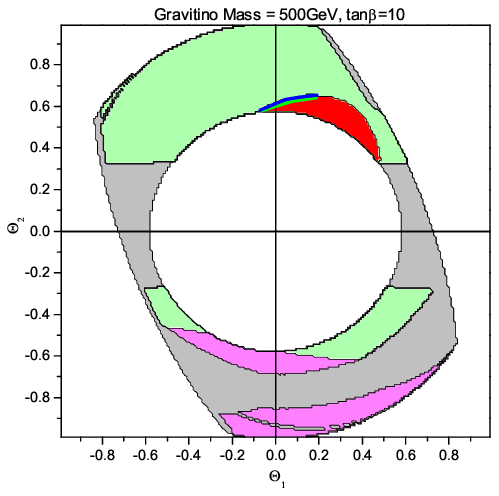}
		\includegraphics[width=0.42\textwidth]{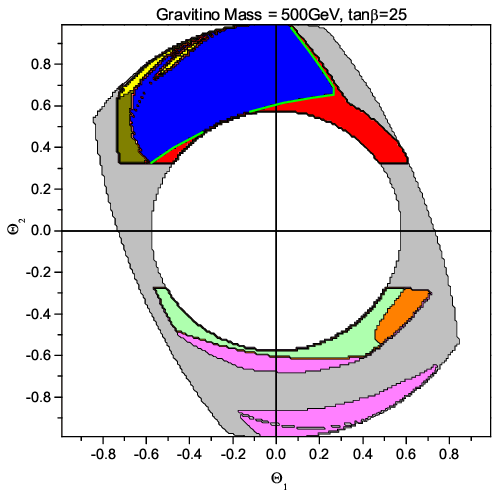}
		\includegraphics[width=0.42\textwidth]{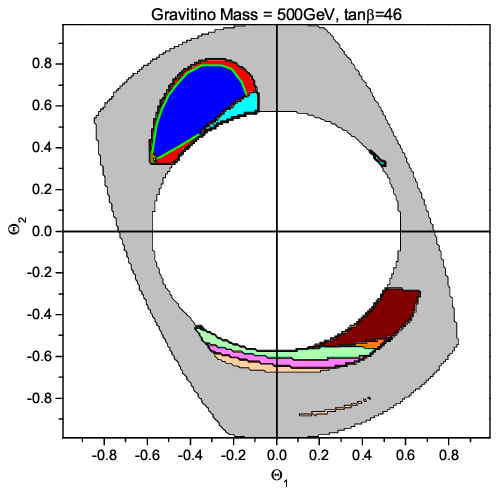}
		\includegraphics[width=0.42\textwidth]{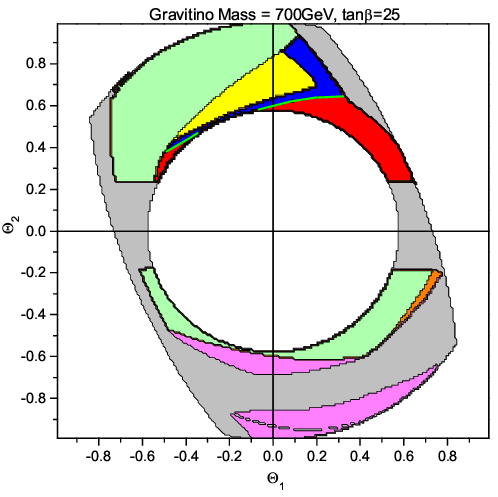}
		\includegraphics[width=0.42\textwidth]{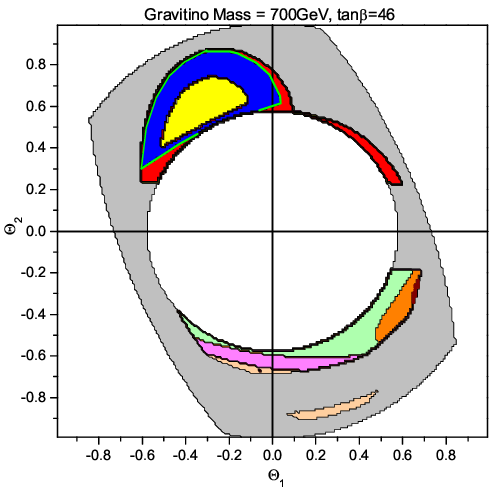}
		\includegraphics[width=0.42\textwidth]{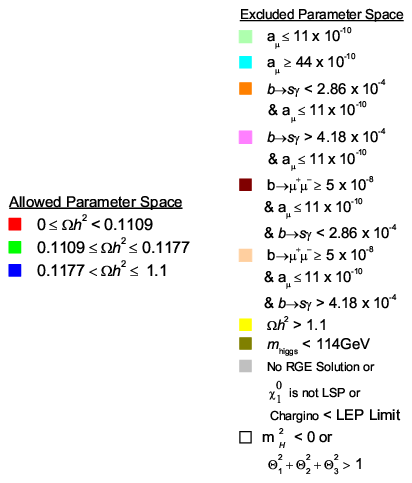}
		\caption{Allowed parameter space for $\textit{u}$-moduli dominated SUSY breaking scenario for an intersecting $D$6-brane model. The five individual charts represent different gravitino masses and tan$\beta$. The chart legend describes the reasons for inclusion and exclusion of the shaded regions.}
	\label{fig:D6_ParamSpace}
\end{figure}

We plot the parameter space in terms of the goldstino angles $\Theta_{1}$ and $\Theta_{2}$ in Fig.~\ref{fig:D6_ParamSpace}. A detailed discussion of the goldstino angles $\Theta_{1}$ and $\Theta_{2}$ and how they relate to the non-universal gaugino masses and scalar masses can be found in~\cite{Chen:2007px}~\cite{Chen:2007zu}. The different shades represent the regions which are allowed or excluded for the reasons noted in the chart legend. These plots focus on the experimental constraints and the dark matter density within the SSC and WMAP regions. Note the small regions excluded by the Higgs mass $m_h$ $<$ 114 GeV satisfy all other constraints, including the SSC dark matter density. All the regions in the allowed parameter space pass the Higgs mass constraint, except as just noted, however, it is not identified on the charts whether the excluded regions meet or fail the higgs mass constraint, though all the excluded regions fail to meet one or more of the remaining constraints. In addition, the regions excluded by $\Omega_{\chi^o} h^{2} >$ 1.1 satisfy all other constraints. The circular region centered at the origin of the plot is excluded for driving $m_{H}^{2}$ to negative values. The region outside the allowed parameter space is excluded since the goldstino angles $\Theta_{1}$, $\Theta_{2}$, and $\Theta_{3}$ do not satisfy the unitary condition $\Theta_{1}^{2} + \Theta_{2}^{2} + \Theta_{3}^{2} = 1$~\cite{Chen:2007px}~\cite{Chen:2007zu}. For more explicit details of the parameter space regions satisfying the experimental constraints, including potential LHC signatures and experimental observables, see~\cite{JMMN}. In the remainder of this work, we focus only on the direct and indirect detection parameters.

\section{WIMP Detection}

Direct detection experiments search for dark matter through an elastic collision of WIMPs with ordinary matter. The lightest neutralino, $\chi_{1}^{0}$, is assumed to be stable, and as such represents the best possible candidate for dark matter, and hence, WIMPs. These WIMPs produce low energy recoils with nuclei. The interaction between the WIMPs and nuclei can be segregated into a spin-independent (SI) part and a spin-dependent(SD) part, where the SI (scalar) interactions are primarily the consequence of elastic collisions with heavy nuclei. First, we consider the SI cross-sections for an intersecting $D$6-brane model, then study the SD interactions.

Both the direct detection cross-sections and the gamma-ray flux are calculated using {\tt MicrOMEGAs 2.1}~\cite{Belanger:2008sj}. For the SI calculation, we use the nucleon form factor coefficient values of

\begin{center}
$f_{d}^{p} = 0.033,~f_{u}^{p} = 0.023,~f_{s}^{p} = 0.26$
\end{center}
\begin{center}
$f_{d}^{n} = 0.042,~f_{u}^{n} = 0.018,~f_{s}^{n} = 0.26$
\end{center}

while for the SD computations, we use the following quark density coefficients

\begin{center}
$\Delta_{u}^{p} = 0.842 \pm 0.012,~\Delta_{d}^{p} = -0.427 \pm 0.013,~\Delta_{s}^{p} = -0.085 \pm 0.018$
\end{center}
\begin{center}
$\Delta_{u}^{n} = \Delta_{d}^{p},~\Delta_{d}^{n} = \Delta_{u}^{p},~\Delta_{s}^{n} = \Delta_{s}^{p}$
\end{center}

In addition, we use $v_{0} = 220$ km/s for the dark matter velocity distribution in the galaxy rest frame, $v_{E} = 244.4$ km/s for the Earth velocity with respect to the galaxy, and $v_{max} = 600$ km/s for the maximal dark matter velocity in the sun's orbit with respect to the galaxy.

\begin{figure}[htp]
	\centering
		\includegraphics[width=0.44\textwidth]{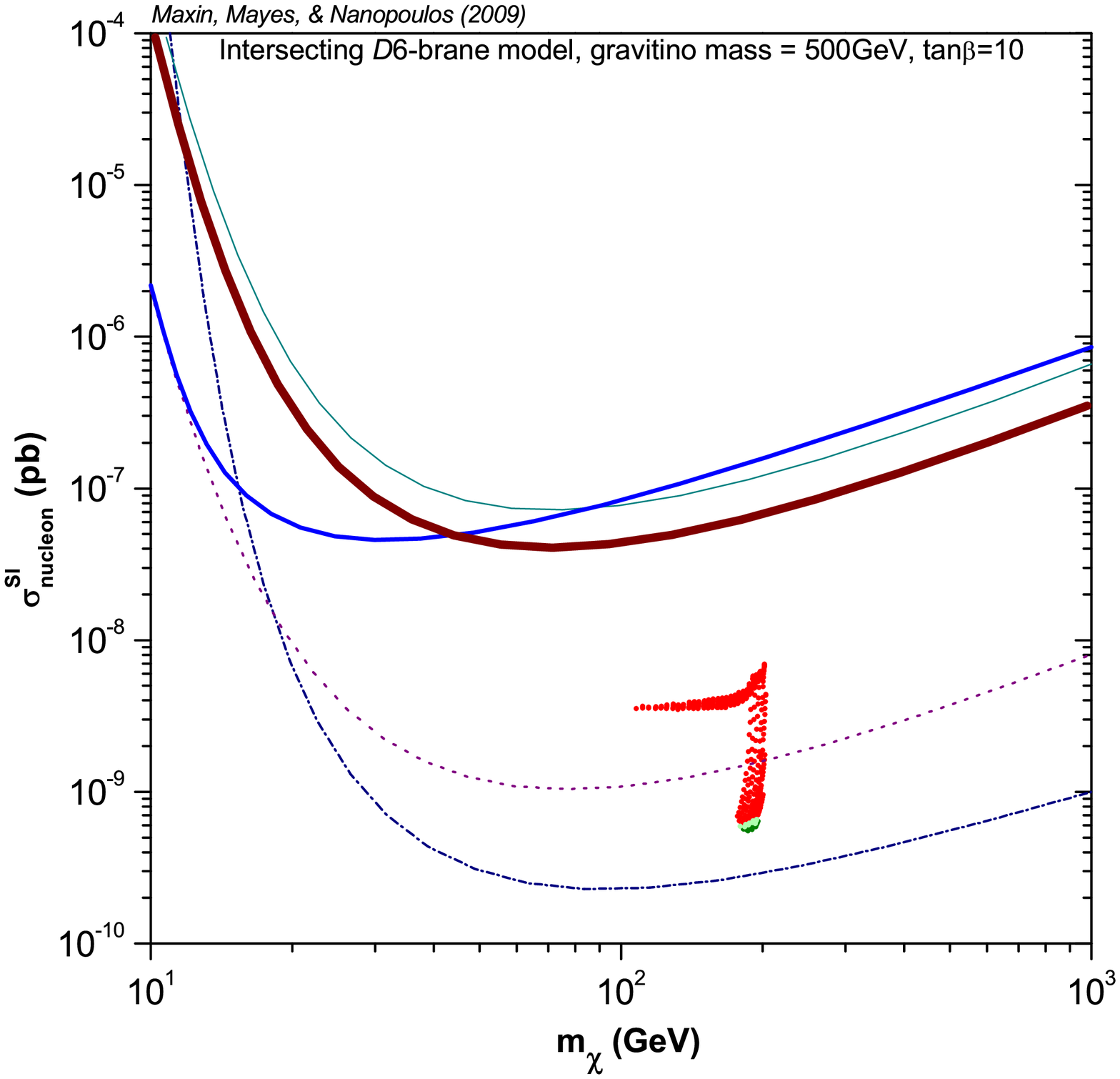}
		\includegraphics[width=0.44\textwidth]{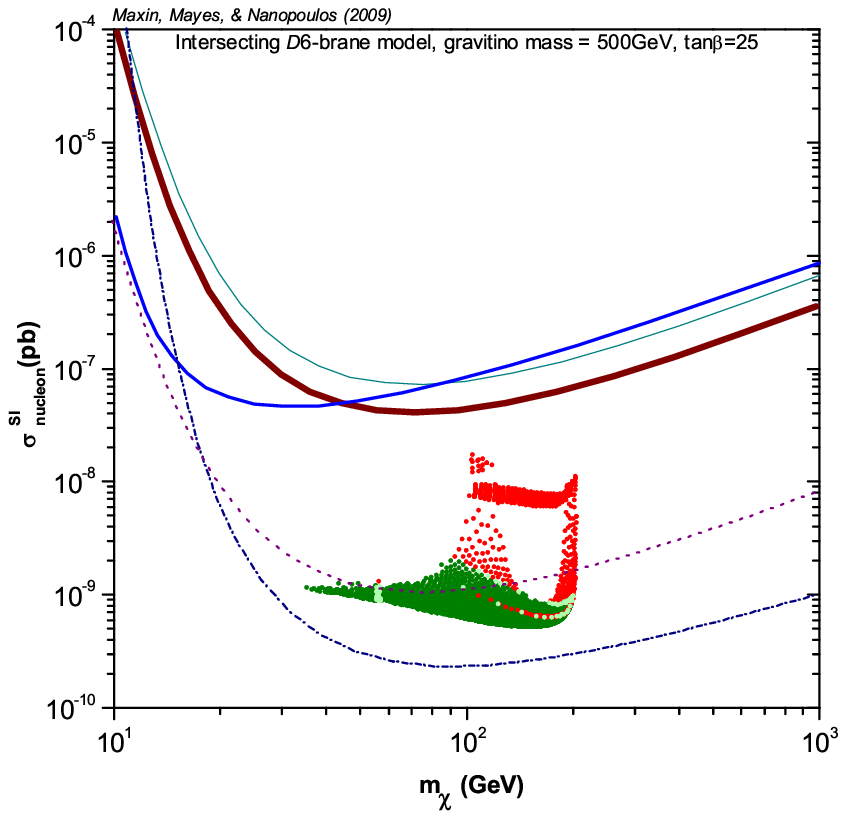}
		\includegraphics[width=0.44\textwidth]{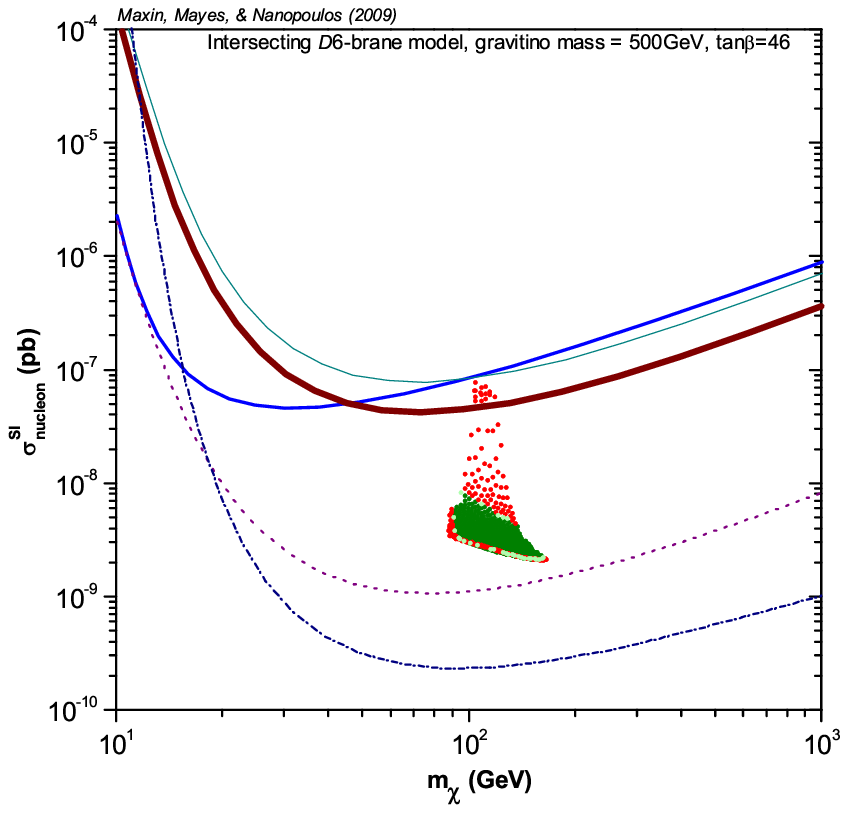}
		\includegraphics[width=0.44\textwidth]{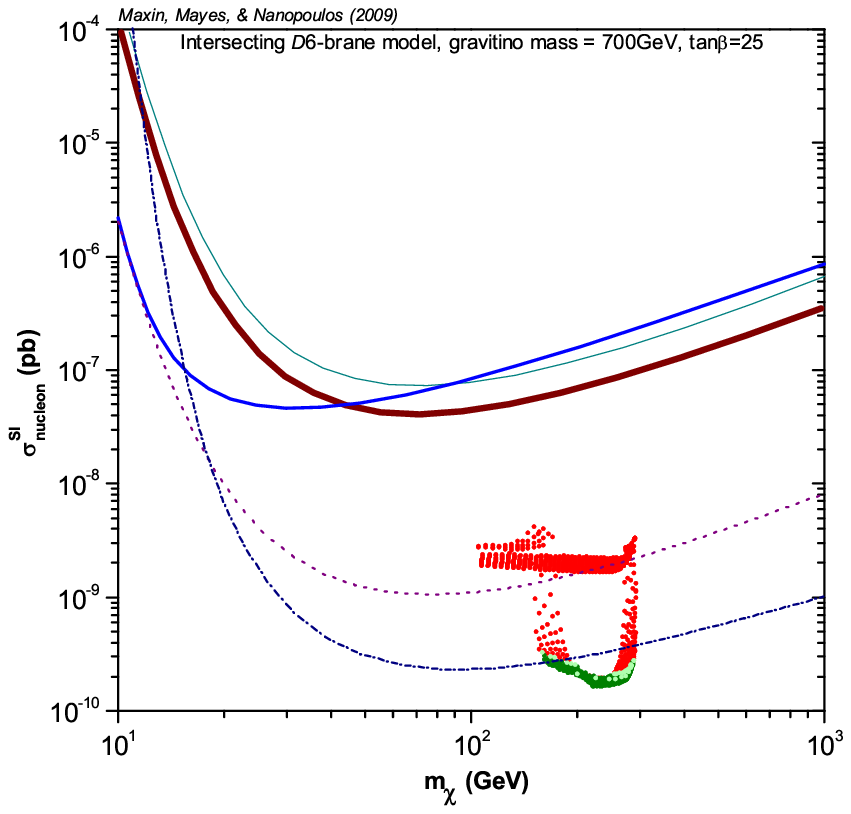}
		\includegraphics[width=0.44\textwidth]{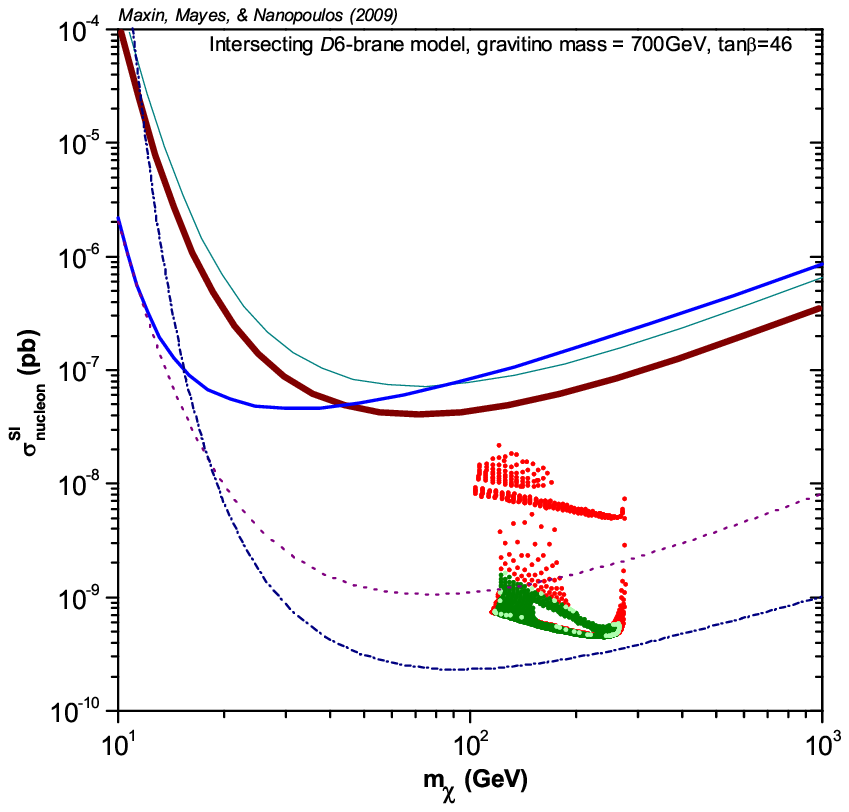}
		\includegraphics[width=0.44\textwidth]{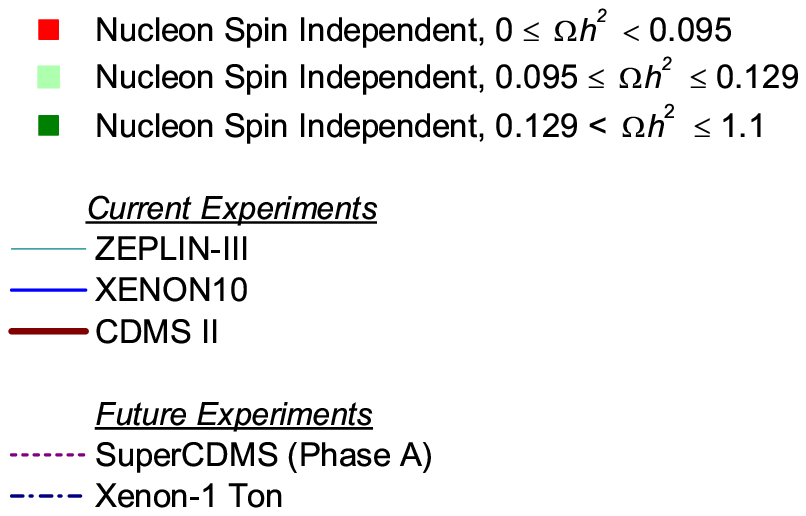}
		\caption{Spin-independent cross-sections of an intersecting $D$6-brane model. Each marker satisfies all experimental constraints for an explicit gravitino mass and tan$\beta$. The three marker colors identify the dark matter density.}
	\label{fig:D6_SpinIndependent}
\end{figure}

In Fig.~\ref{fig:D6_SpinIndependent}, we plot the SI cross-sections for an intersecting $D$6-brane model. The cross-sections and flux were calculated only for those regions of the parameter space satisfying all the experimental constraints. Those allowed regions are shown in Fig.~\ref{fig:D6_ParamSpace}. The plots in Fig.~\ref{fig:D6_SpinIndependent} are subdivided by dark matter density, where for clarity we use the 2$\sigma$ WMAP limits. The most recent experimental results for Zeplin-III~\cite{Lebedenko:2008gb}, Xenon 10~\cite{Angle:2007uj}, and CDMS II~\cite{Ahmed:2008eu} are shown, in addition to the projected sensitivity of the future SuperCDMS~\cite{Schnee:2005pj} and Xenon-1 Ton~\cite{Aprile:2005mz} experiments. Only for $m_{3/2}$ = 500 GeV and tan$\beta$ = 10 are the cross-sections within the current experimental limits, however, in this case there is only a small region of the allowed parameter space within the latest CDMS results, where these points have a very small dark matter density only allowed since we removed the lower WMAP 2$\sigma$ boundary. Most of the points will be within the experimental limits of the SuperCDMS and Xenon-1 Ton future experiments, potentially providing incentive for the design and development of the next generation of dark matter direct detection experiments. In the SSC region, we allow for a dilution factor of $\cal{O}$(10), resulting in a dark matter density up to $\Omega_{\chi^o} h^{2} \sim 1.1$, permitting the inclusion of more points. As can be seen in Fig.~\ref{fig:D6_SpinIndependent}, in general, the SSC regions have a smaller cross-section than the WMAP regions. The dark matter density $\Omega_{\chi^o} h^{2}$ is inversely proportional to the annihilation cross-section $\left\langle \sigma_{\textit{ann}}\textit{v}\right\rangle$, so one expects the points with a higher $\Omega_{\chi^o} h^{2}$ to possess a smaller annihilation cross-section, as depicted in Fig.~\ref{fig:D6_SpinIndependent}.

\begin{figure}[htp]
	\centering
		\includegraphics[width=0.44\textwidth]{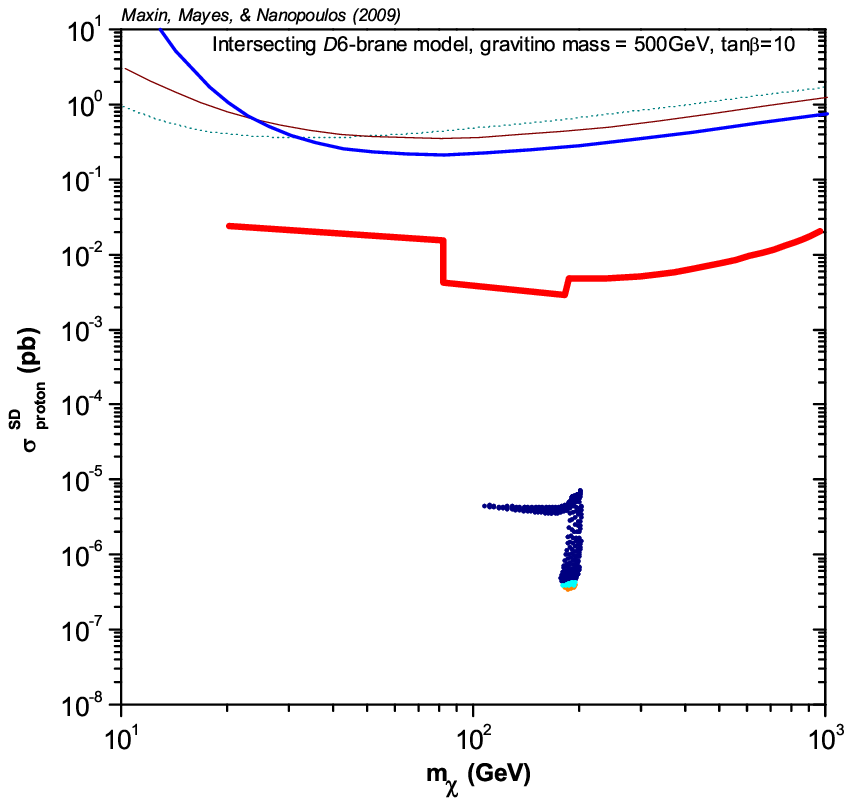}
		\includegraphics[width=0.44\textwidth]{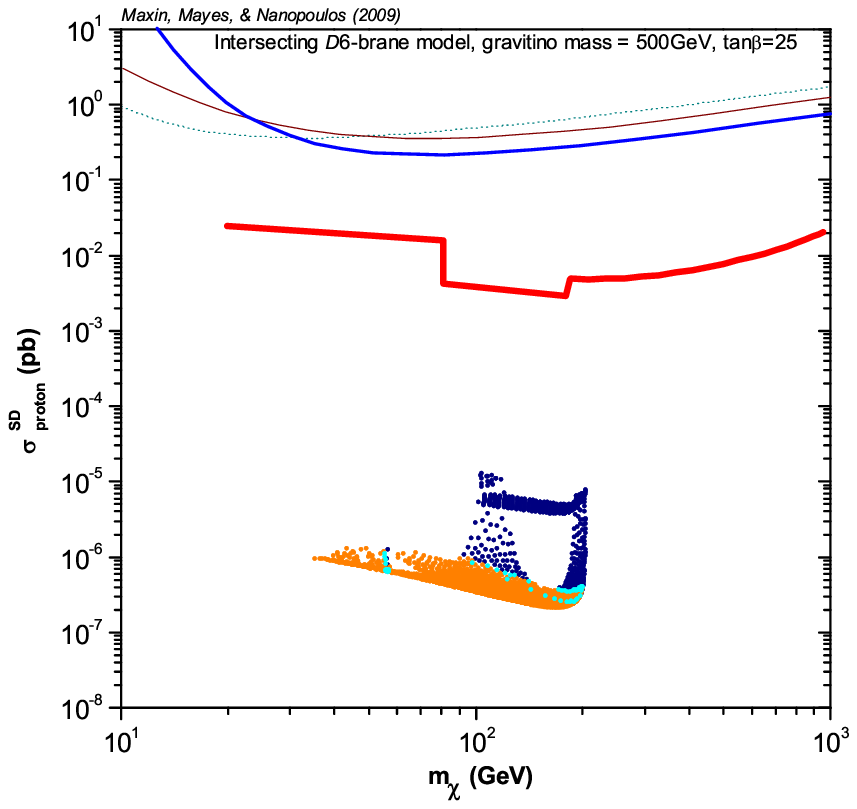}
		\includegraphics[width=0.44\textwidth]{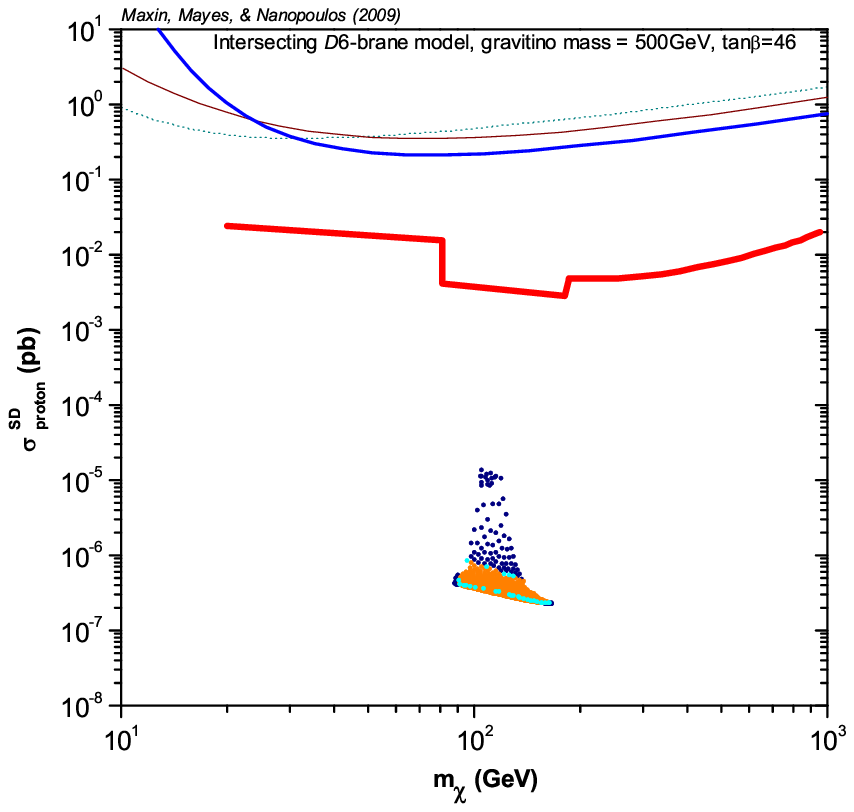}
		\includegraphics[width=0.44\textwidth]{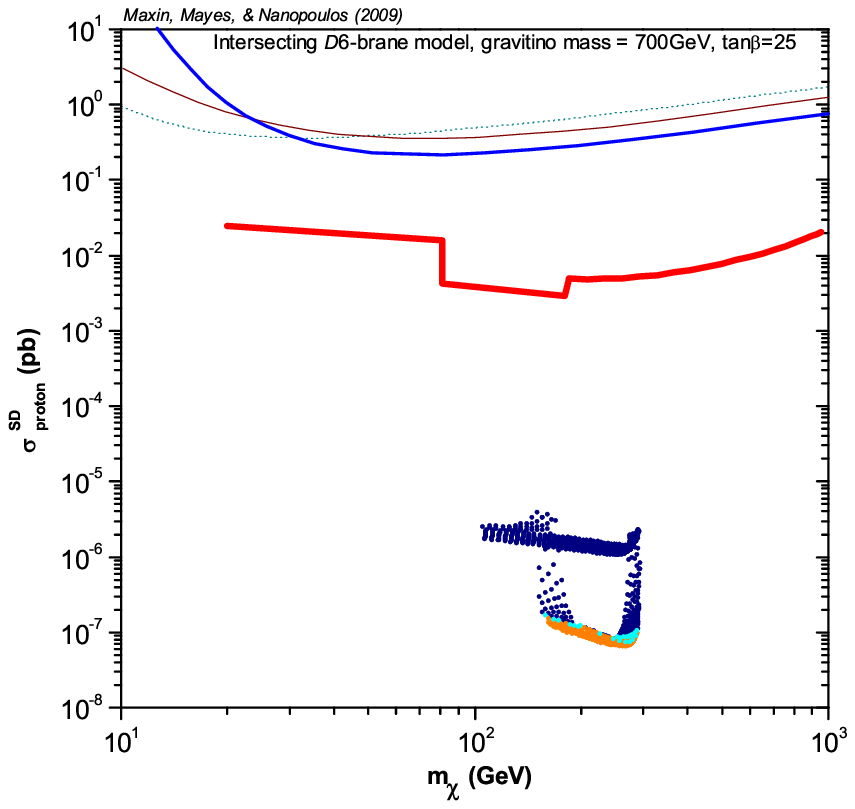}
		\includegraphics[width=0.44\textwidth]{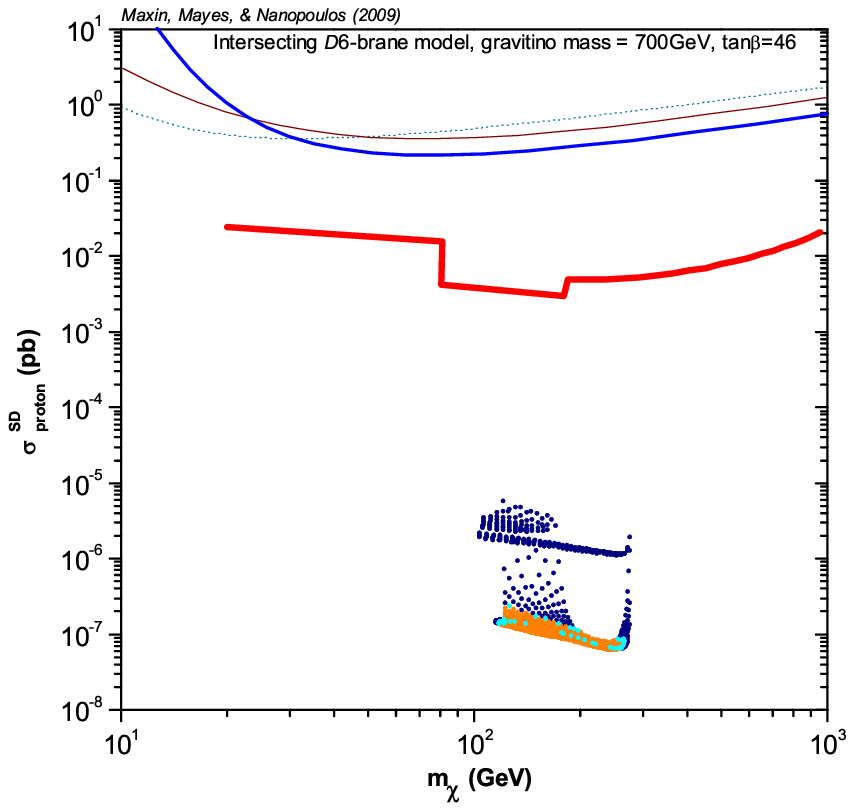}
		\includegraphics[width=0.44\textwidth]{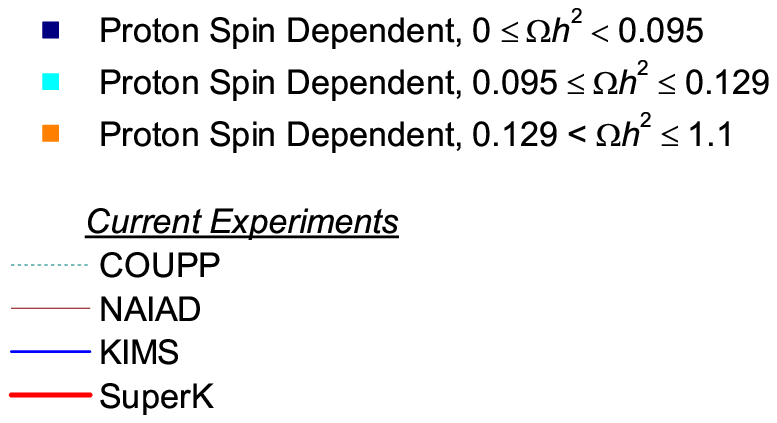}
		\caption{Proton spin-dependent cross-sections of an intersecting $D$6-brane model. Each marker satisfies all experimental constraints for an explicit gravitino mass and tan$\beta$. The three marker colors identify the dark matter density.}
	\label{fig:D6_SpinDependent}
\end{figure}

The proton SD cross-sections are shown in Fig.~\ref{fig:D6_SpinDependent}. The format of the SD charts is similar to the SI charts. For comparison of the intersecting $D$6-brane model cross-sections to the current experimental limits, we show the latest results for COUPP~\cite{COUP}, NAIAD~\cite{NAIA}, KIMS~\cite{Lee.:2007qn}, and SuperK~\cite{Desai:2004pq}. We also calculated the neutron SD cross-sections (not shown), though there was only a slight difference between the proton and neutron SD. The patterns were generally the same, but the neutron SD cross-sections were slightly larger, and the shape of the SD patterns is essentially identical to the SI patterns. None of the intersecting $D$6-brane model points are within the current experimental limits of the SD dark matter detectors, and in fact, they are still three orders of magnitude away from the discovery region. Again, since $\Omega_{\chi^o} h^{2} \sim \frac{1}{\left\langle \sigma_{\textit{ann}}\textit{v}\right\rangle}$, we see in Fig.~\ref{fig:D6_SpinDependent} that the SSC points have in general a smaller annihilation cross-section than the WMAP points.

\section{Indirect Dark Matter Detection}

\begin{figure}[htp]
	\centering
		\includegraphics[width=0.44\textwidth]{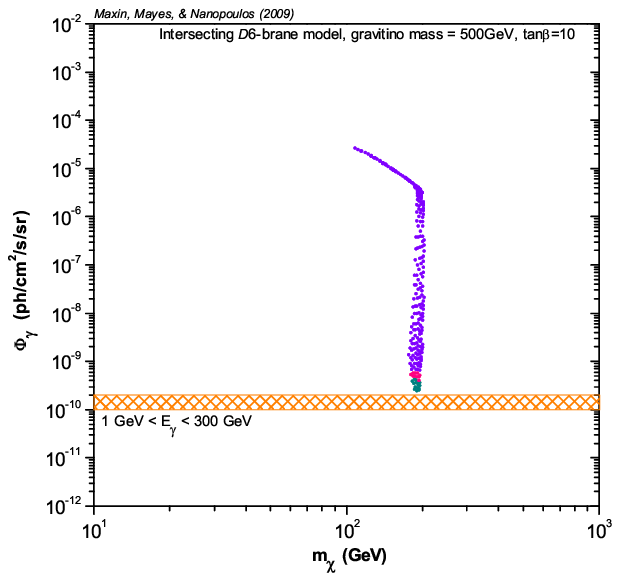}
		\includegraphics[width=0.44\textwidth]{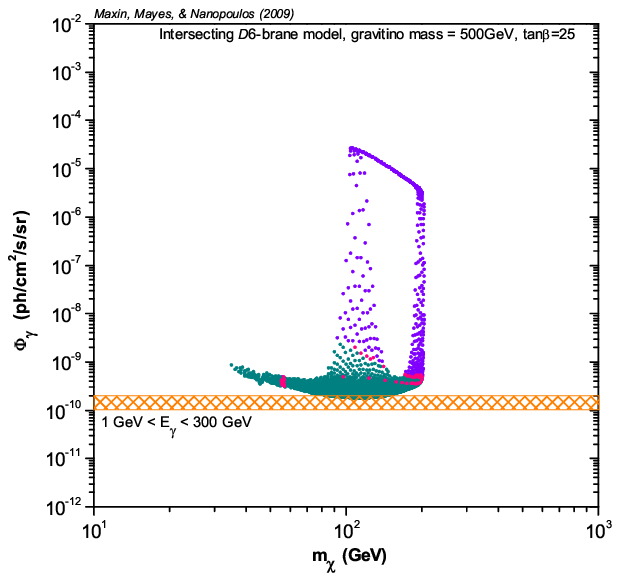}
		\includegraphics[width=0.44\textwidth]{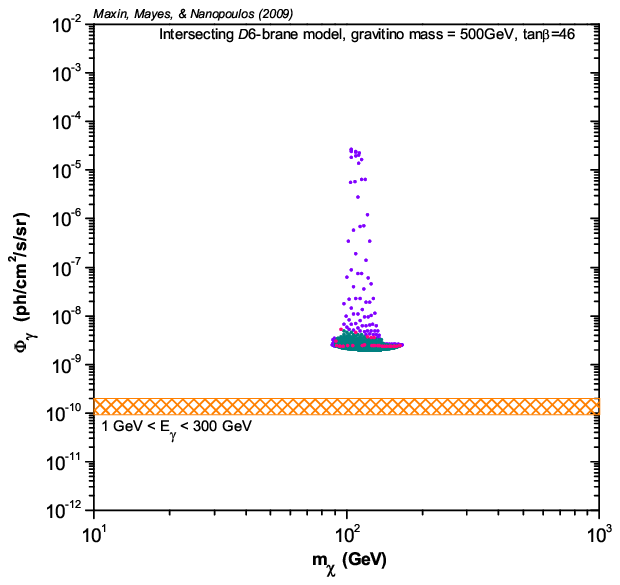}
		\includegraphics[width=0.44\textwidth]{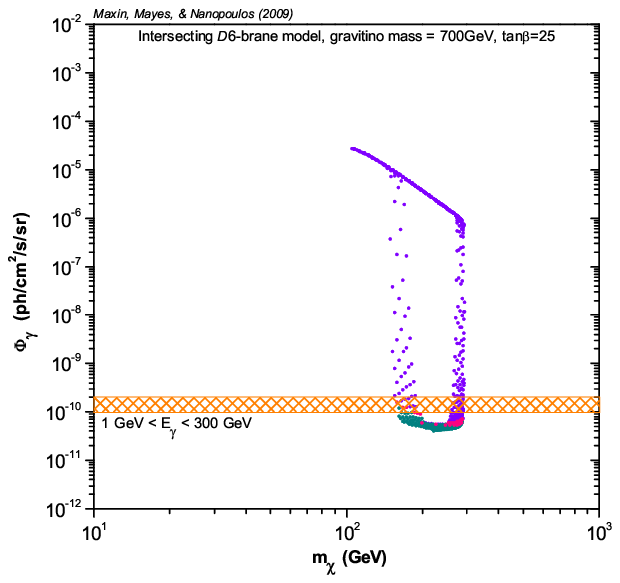}
		\includegraphics[width=0.44\textwidth]{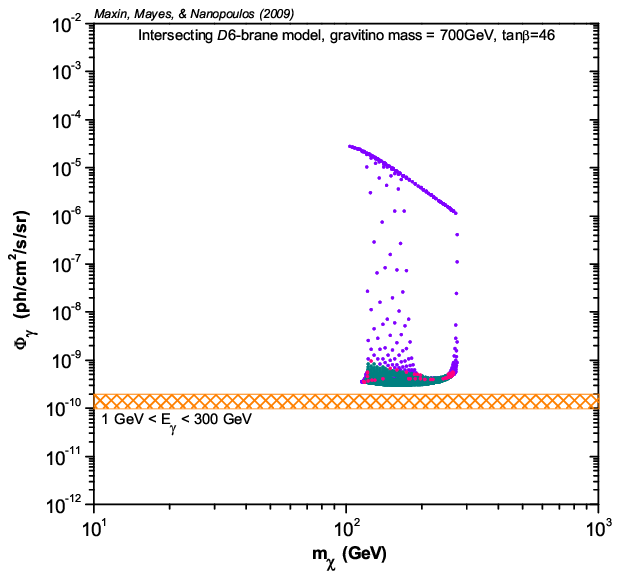}
		\includegraphics[width=0.44\textwidth]{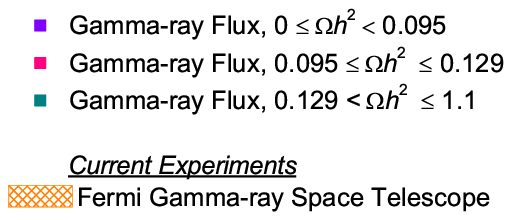}
		\caption{Gamma-ray flux of an intersecting $D$6-brane model. Each marker satisfies all experimental constraints for an explicit gravitino mass and tan$\beta$. The three marker colors identify the dark matter density.}
	\label{fig:D6_GammaFlux}
\end{figure}

Indirect detection experiments search for high energy neutrinos, gamma-rays, positrons, and anti-protons emanating from neutralino annihilation in the galactic halo and core, or in the case of neutrinos, in the core of the sun or the earth. In this work, we focus only on the flux of gamma-rays $\Phi_{\gamma}$ in the galactic core or halo. The gamma-ray flux $\Phi_{\gamma}$ for the intersecting $D$6-brane model is shown in Fig.~\ref{fig:D6_GammaFlux}, including the projected sensitivity of the Fermi experiment~\cite{Morselli:2002nw}. The sensitivity is not constant, but is a function of photon energy, and for this reason, to be precise, we delineate it using a band. Most of the points allowed by the experimental constraints will be within the sensitivity of the Fermi telescope. As mentioned in the Introduction, two possible decay channels where WIMPs can produce gamma-rays in the galactic core and halo are $\widetilde{\chi}_{1}^{0} \widetilde{\chi}^{0}_{1} \rightarrow \gamma \gamma$ and $\widetilde{\chi}_{1}^{0} \widetilde{\chi}_{1}^{0} \rightarrow q \overline{q} \rightarrow \pi^{0} \rightarrow \gamma\gamma$. Hence, the flux of gamma-rays is directly dependent upon the annihilation cross-section. Fig.~\ref{fig:D6_SpinIndependent} and Fig.~\ref{fig:D6_SpinDependent} show and we have explained that the SSC points have a smaller annihilation cross-section. Consequently, we expect the SSC points to also exhibit a smaller gamma-ray flux $\Phi_{\gamma}$, and accordingly, this is illustrated in Fig.~\ref{fig:D6_GammaFlux}.

\begin{figure}[htp]
	\centering
		\includegraphics[width=0.44\textwidth]{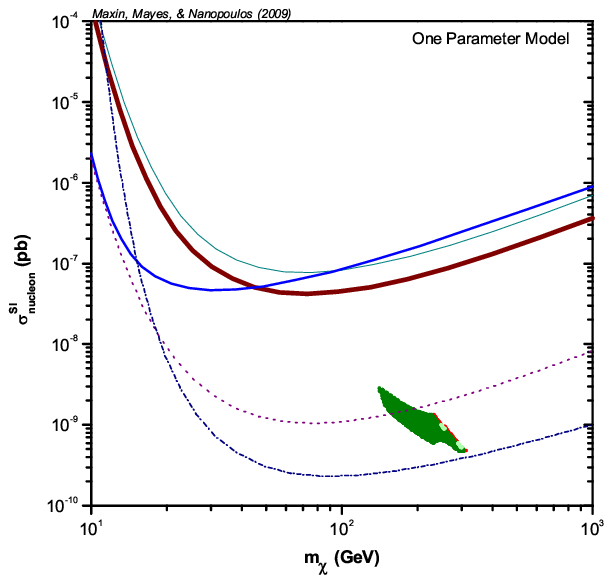}
		\includegraphics[width=0.44\textwidth]{D6_SI6a.eps}
		\includegraphics[width=0.44\textwidth]{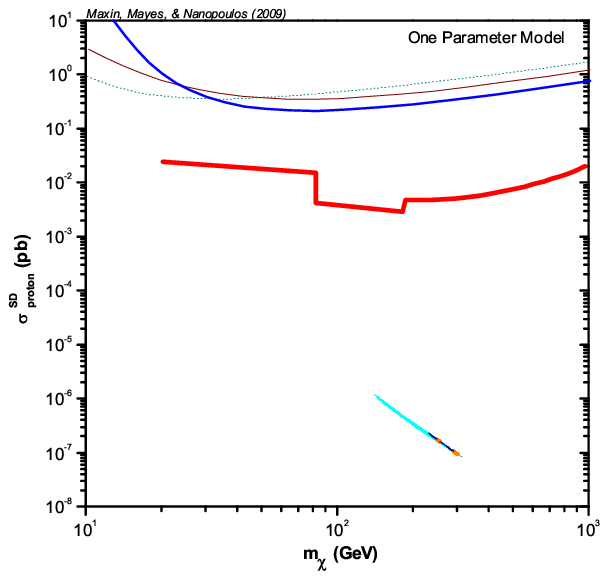}
		\includegraphics[width=0.44\textwidth]{D6_SD6a.eps}
		\includegraphics[width=0.44\textwidth]{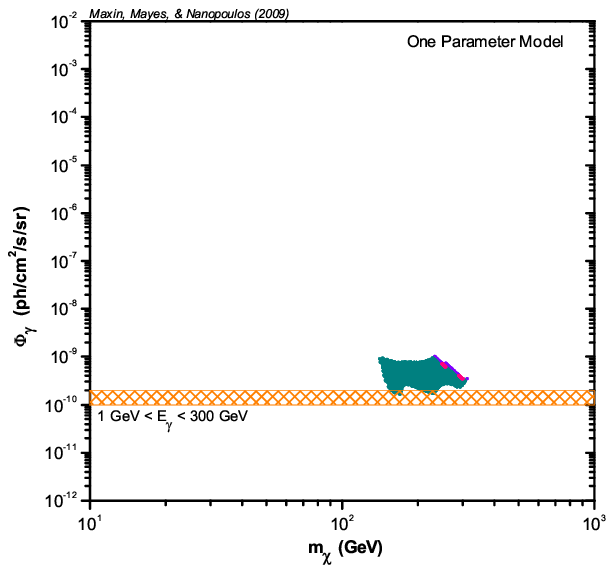}
		\includegraphics[width=0.44\textwidth]{D6_ID6a.eps}
		\caption{Spin-independent cross-section, proton spin-dependent cross-section, and gamma-ray flux for the one-parameter model. Each marker satisfies all experimental constraints. The different marker colors identify the dark matter density.}
	\label{fig:OPM}
\end{figure}

It is an intriguing question as to how a model with non-universal soft-supersymmetry breaking terms, such as an intersecting $D$6-brane model, compares to a model with universal soft-supersymmetry breaking terms, for example, mSUGRA. The one-parameter model (OPM)~\cite{Lopez:1993rm,Lopez:1994fz,Lopez:1995hg,Maxin:2008kp} is a highly constrained small subset of mSUGRA, where all the soft-supersymmetry breaking terms may be fixed in terms of the gaugino mass $m_{1/2}$. The OPM has universal soft-supersymmetry breaking terms, so it is ideal to compare to the $D$6-brane model. Details of the phenomenology of the OPM using the most recent measurements of the experimental constraints can be found in~\cite{Maxin:2008kp}. The parameter space of the OPM is quite constrained by the experimental constraints, and this leads to small regions of allowed direct and indirect detection parameters. In Fig.~\ref{fig:OPM} we plot the direct and indirect detection parameters of the OPM. As described in~\cite{Maxin:2008kp}, the range of tan$\beta$ for spectra that satisfy all the experimental constraints in the WMAP region is 35.2 $<$ tan$\beta$ $<$ 38, while the range in the SSC region is 10.2 $<$ tan$\beta$ $<$ 38. Note that the points shown in Fig.~\ref{fig:OPM} are for all tan$\beta$ within the aforementioned ranges. However, it can be concluded from Fig.~\ref{fig:OPM} that the points with the same WIMP mass do exhibit the same characteristics as the points in the intersecting $D$6-brane model. For the same WIMP mass, the WMAP spectra have a larger annihilation cross-section, and hence, gamma-ray flux than the SSC points, due to the fact that in the WMAP region, we are not allowing for the $\cal{O}$(10) dilution factor to $\Omega_{\chi^o} h^{2}$.

\section{Conclusion}

Much advancement has been made in the last few years toward the discovery of dark matter. Current generation direct detection experiments that search for elastic collisions of WIMPs off nuclei have come within shouting distance of the allowed parameter space of models with universal soft-supersymmetry breaking terms such as mSUGRA. Furthermore, the Fermi Gamma-ray space telescope is edging closer to the parameter space of these same models. In light of this experimental progress, it is a good time to start examining the direct and indirect detection parameters of semi-realistic string models. To this end, we began an investigation of the experimental detection parameters for a particular string-derived model with many appealing phenomenological properties. There are various theoretical models currently offered, so our goal is present the phenomenology of a promising new model, in contrast to the usual standard, mSUGRA. In this work, we investigated an intersecting $D$6-brane model on a Type IIA orientifold that overcomes the persistent problems experienced by many Type II string vacua, namely that of gauge coupling unification and the generation of masses for the first two generations of quarks and leptons. This model exhibits automatic gauge coupling unification and allows the correct masses and mixings for all quarks and leptons to be obtained. As a consequence, we presented the spin-independent and proton spin-dependent cross-sections. We find that only a small region of the allowed parameter space is within the current limits of the direct detection experiments. Regions with a larger $\Omega_{\chi^o} h^{2}$ have smaller cross-sections, thus cross-sections for SSC are smaller than those of WMAP. Additionally, we illustrated the galactic gamma-ray flux for this model resulting from neutralino annihilations. We discover that most of the regions of the $D$6-brane model allowed parameter space will be within the sensitivity of the Fermi telescope. 

\section{Acknowledgments}

This research was supported in part by the Mitchell-Heep Chair in High Energy Physics (CMC), by the Cambridge-Mitchell Collaboration in Theoretical Cosmology, and by the DOE grant DE-FG03-95-Er-40917.

\end{document}